\documentclass[referee]{cjaa}           

\usepackage{graphicx}                   
\input{epsf.sty}                        
\input{psfig.sty}                       

\newcommand{\Mg}{Mg{\sc ii} }
\newcommand{\eMg}{\rm Mg {\scriptstyle II} }
\newcommand{\C}{\rm C{\sc iv} }
\newcommand{\eC}{C{\scriptstyle IV} }
\newcommand{\Hb}{\rm H$\beta$ }
\newcommand{\RcMg}{R_{\rm BLR}\,-\,\lambda\,L_{\lambda}(3000\AA)}
\newcommand{\RMg}{R_{\rm BLR}\,-\,L_{\eMg}}
\newcommand{\RcC}{R_{\rm BLR}\,-\,\lambda\,L_{\lambda}(1350\AA)}

\newcommand{\RC}{R_{\rm BLR}\,-\,L_{\eC}}

\begin{document}

   \title{Estimate black hole masses of AGNs using ultraviolet emission line properties}

   \volnopage{Vol.0 (200x) No.0, 000--000} 
   \setcounter{page}{1} 

   \author{Min-Zhi Kong
      \inst{1,2}\mailto{}
   \and  Xue-Bing Wu
      \inst{2}
   \and  Ran Wang
      \inst{2}
   \and  Jin-Lin Han
      \inst{1}
      }
   \offprints{Min-Zhi Kong}                   

   \institute{National Astronomical Observatories, Chinese Academy of Sciences, Beijing 100012, China\\
            \email{kmz@bao.ac.cn}
        \and
            Department of Astronomy, Peking University, Beijing 100871, China\\
          }
   \date{Received; accepted}

   \abstract{Based on the measured sizes of broad line region of the
reverberation-mapping AGN sample, two new empirical relations are
introduced to estimate the central black hole masses of radio-loud
high-redshift ($z\,>\,0.5$) AGNs. First, using the archival $IUE/HST$
spectroscopy data at UV band for the reverberation-mapping objects, we
obtained two new empirical relations between the BLR size and \Mg/\C
emission line luminosity. Secondly, using the newly determined black
hole masses of the reverberation-mapping sample for calibration, two
new relationships for determination of black hole mass with the full
width of half maximum and the luminosity of \Mg/\C line are also
found. We then apply the relations to estimate the black hole masses
of AGNs in Large Bright Quasar Surveyq and a sample of radio-loud
quasars. For the objects with small radio-loudness, the black hole
mass estimated using the $R_{\rm BLR} - L_{\eMg/\eC}$ relation is
consistent with that from the $R_{BLR}\,-\,L_{3000\AA/1350\,\AA}$
relation. But for radio-loud AGNs, the mass estimated from the
$R_{BLR}\,-\,L_{\eMg/\eC}$ relation is systematically lower than that
from the continuum luminosity $L_{3000\AA/1350\AA}$. Because jets
could have significant contributions to the UV/optical continuum
luminosity of radio-loud AGNs, we emphasized again that for radio-loud
AGNs, the emission line luminosity may be a better tracer of the
ionizing luminosity than the continuum luminosity, so that the
relations between the BLR size and UV emission line luminosity should
be used to estimate the black hole masses of high redshift radio-loud
AGNs.  }

\authorrunning{M.Z. Kong, X.B. Wu, R.~W, J.L. Han}
\titlerunning{Estimate black hole masses of AGNs}

\maketitle
\section{Introduction}
\label{sect:intro}
Black hole masses of high-redshift AGNs ($z\ga1$) are essential to
understand the early history of the universe and the formation of
supermassive black holes. The mass of black hole can be estimated by
$M_{BH}\,\sim\, R_{BLR}\,V^{2}/{G}$ assume that gas around a black
hole is virialized.  Here $V$ is the characteristic velocity of BLR
gas at distance $R_{BLR}$ from the center of a AGN (Peterson \& Wandel
1999; Peterson 1993), $G$ is the gravitational constant.  The velocity
can be estimated from $V_{FWHM}$ (full width at half maximum) of
emission lines. For randomly distributed BLR clouds, $V\,=(\sqrt{3}/2)
V_{FWHM}.$ The exact scale factor between $V$ and $V_{FWHM}$ depends
on the structure, kinematics, and orientation of BLR (Peterson \&
Wandel 1999, 2000; McLure \& Dunlop 2001; Wu \& Han et al. 2001; Zhang
\& Wu, 2002). Taking practical parameters, the virial mass of a black
hole can thus be expressed in the form of (see Kaspi et al. 2000)
$$
   M_{BH}\,=\,1.464\times10^{5}\left(\rm \frac
   {R_{BLR}}{lt-days}\right)\,\left(\frac {
   V_{\rm FWHM}}{10^{3}\,\rm km\,s^{-1}}\right)^{2}\,M_{\sun}.
$$

The reverberation mapping technique is the most important method to
study the geometry and kinematics of gas in BLR (Blandford \& McKee
1982; Peterson 1993, 1999, 2000; Horne et al. 2004). The BLR size,
$R_{BLR}$, can be deduced from time delay of luminosity variations of
broad emission lines, often  \Hb  emission line (see Peterson et
al. 2004 and references therein), relative to that of ionizing
continuum. In principle, this technique can be applied to any AGN. In
practice, monitor observations of AGNs for the time delay are very
time-consuming as it can have a time scale of weeks, months or even
years. Up to now, only 20 Seyfert~1 galaxies and 17 nearby quasars
(Kaspi et al. 2000; Wandel et al. 1999; Peterson et al. 1998;
Santos-Lleo et al. 2001) have been well monitored in the
reverberation-mapping studies.

To ease the problem, Kaspi et al. (2000) first successfully obtained
an empirical relation between BLR sizes of AGNs in the reverberation
mapping sample and optical continuum luminosites at $5100\,\AA$,
i.e. the $R_{BLR}\,-\,\lambda\,L_{\lambda}(5100\AA)$ relation, which
can be used to estimate BLR size. Peterson et al. (2004) recently have
re-analyzed the reverberation mapping data for 35 AGNs (after PG
1351+640 and PG 1704+608 were omitted) and obtained the improved time
delays and hence BLR sizes and black home masses.  Using the improved
BLR sizes in Peterson et al.(2004) and the new cosmological model,
Kaspi et al. (2005) recently investigated the relation between BLR
sizes and luminosities of  \Hb  line and continuum at $5100\AA$,
$1450\AA$, $1350\AA$ and 2-10~keV.  Specifically, they found $
R_{BLR}\,\propto \lambda\,L_{\lambda}(5100\AA)^{0.67\,\pm\,0.05} $
with about 40\% intrinsic scatter.
The mass of black hole of a low-redshift ($z \la 0.8$) AGN can thus be
easily estimated using the above relation on $R_{BLR}$ and $V_{FWHM}$
(Laor 2000; McLure \& Dunlop et al. 2001; Wandel 2002; Wu \& Liu
2004).  For objects at higher redshift ($0.8 \la z \la 2.5$), as
McLure \& Jarvis (2002) stated, $V_{FWHM}$ of \Mg can be adopted to
substitute for $V_{FWHM}$ of  \Hb , mainly because they both are
strong, fully permitted and low-ionization lines with similar
ionization potentials and are emitted at approximately the same radius
from the central ionizing source. It is not surprising that the FWHM
values for \Mg and  \Hb  are very closely related for reverberation
mapping objects. McLure \& Jarvis (2002) obtained a relation between
the $R_{BLR}$ from  \Hb  time lag and the continuum luminosity
$\lambda\,L_{\lambda}(3000\AA)$.  Note, however, a spectrum near
$3000\AA$ is seriously contaminated by the blended  Fe{\sc ii}  and
the Balmer continuum, so it is important to remove their effects when
measuring the FWHM of \Mg. The  Fe{\sc ii}  emission can be
removed from the continuum by model-fitting (McLure \& Dunlop 2004),
while the Balmer continuum contribution is not obvious in the residual
spectrum and difficult to remove. For two AGN samples, LBQS (Forster
et al. 2001) and the Molonglo quasar sample (Kapahi et al. 1998),
McLure \& Jarvis (2002) calculated the black hole masses using the two
relations, $R_{BLR}\,-\,\lambda\,L_{\lambda}(5100\AA)$ and
$R_{BLR}\,-\,\lambda\,L_{\lambda}(3000\AA)$, and found that results
agree well with each other.

For a given object, NGC 5548, Peterson \& Wandel (1999) found from
time delay of different emission lines that the same virial
relationship exists for lines emitted at different distances from the
central black hole, such as  \Hb , \C$\lambda1549$,
He{\sc ii} $\lambda1640$. Similar results have been obtained also for
NGC 7469 and 3C 390.3 (Peterson \& Wandel 2000). This suggests a
possibility to determine the black hole mass using the line properties
and/or continuum luminosity at short wavelength around
\C $\lambda1549$ (Vestergaard 2002).  To estimate the velocity, as
Vestergaard (2002) pointed out, the FWHM of \C  can be used because
the line is available for large redshift range between $z\sim 1-5$ and
not much affected by strong absorption lines and the  Fe{\sc ii}
emission. To estimate the $R_{BLR}$, Kaspi et al. (2005) has got the
$R_{BLR}\,-\,\lambda\,L_{\lambda}(1350\AA)$ and
$R_{BLR}\,-\,\lambda\,L_{\lambda}(1450\AA)$ relations with a power
index about $0.55\pm0.05$. Using the $R_{BLR}(H_\beta)-L$ relation and
the FWHM of \C , one can estimate black hole mass. The mass estimates
can be calibrated using the reverberation mapping results because the
BLR size of \C  and  \Hb  are probably different.

In most cases, the observed continuum luminosity of an AGN is mainly
contributed from its nucleus. But some fraction of the continuum comes
from the nonthermal emission of jet and host galaxy. Especially for
radio-loud quasars and BL Lac objects, jets could contribute
significantly to the continuum radiation. Jet emission has so far been
detected in all kinds of radio sources at
optical/UV/X-ray/$\gamma$-ray band (Jester 2003). More than ten
UV/optical jets have been found (O'Dea, et al. 1999; Scarpa \& Urry
2002; Parma et al. 2003; Scarpa et al. 1999). To diminish the jet
contribution, as well as the continuum radiation from host galaxies
(though it is much weaker than the AGN), the line luminosity should be
used to deduce the $R_{BLR}\,-\,L$ relation.  Wu et al. (2004) have
suggested a new relation between the BLR size and the  \Hb  emission
line luminosity, and have shown that, for radio-loud AGNs, black hole
masses estimated using such a new relation are systematically lower
than those derived using the $\rm
R_{BLR}\,-\,\lambda\,L_{\lambda}(5100\AA)$ relation. We noticed that
the relations for $\rm R_{BLR}$ with the continuum luminosity, as
obtained in Kaspi et al. (2005), were deduced from the 35
reverberation mapping AGNs. Most of them are radio-quiet objects and
only five of them, including PG~1704+608, with radio-loudness greater
than 10 and only one object, PG 1226+023 (3C 273), have radio-loudness
greater than 1000 (see Nelson 2000). Therefore the presence of a few
radio-loud objects only has minor effect to the $R - L$
relationships. The $R - L$ relation derived from the reverberation
mapping AGN sample therefore is not the best to estimate black hole
masses of radio-loud objects for the reasons we mentioned above.

Similar to that of  \Hb  line, the \Mg, \C  line luminosities may
be the better tracers to the ionizing luminosity for radio-loud AGNs
than the UV continuum luminosities. Although the detailed line
radiation mechanisms are different, with  \Hb  being a recombination
line and \Mg, \C  being collisionally excited lines, these three
lines are all permitted lines produced by the photo-ionization
process. Therefore we expect that the luminosities of these lines can
all trace the ionizing luminosity but with different calibration
factors.  In fact, we have investigated the correlations between the \Mg,
\C  line luminosities and the continuum luminosities or  \Hb  line
luminosity for the reverberation mapping AGNs, and found that the
correlation coefficients are all greater than 0.9.

The main purpose of this work is to find relations between the BLR
size and the \Mg, \C  emission line luminosity which can be used to
estimate black hole mass for high-redshift radio-loud AGNs.  We will
compare the black hole masses obtained using, respectively, the
relation between the BLR size and the UV line luminosity, and that
between the BLR size and the UV continuum luminosity, and then examine
whether the black hole masses for radio-loud high-redshift objects are
systematically overestimated using the UV continuum luminosity.

In Sect.2, we present the spectral measurements of the \Mg, \C
emission lines obtained from archival UV data of $IUE/HST$
observations for the reverberation-mapping AGNs. In Sect.3, we
investigate the relations between the BLR size and \Mg, \C
emission line luminosity using data for the reverberation-mapping
AGNs. In Sect.4, we apply the relations to other AGN samples for black
hole mass estimation, and compare the black hole masses obtained by
using different relations. In Sect.5, we present our conclusions and
discussions.

Throughout the paper, a cosmological model, with $H_{0} = 70~\rm
km\,s^{-1}\,Mpc^{-1}$,\,\, $\Omega_{\Lambda} = 0.7$,\,\, $\rm
\Omega_{M} = 0.3$, has been adopted.

\section{Data for the reverberation-mapping AGNs}

To investigate possible relations between the BLR size and \Mg, \C
emission line luminosity, we collect data for reverberation-mapping
AGNs.  We take the average BLR size values for 35 AGNs in Peterson et
al. (2004) obtained from the time delay of  \Hb  line. There is no
systematically measurements of the luminosities of \Mg, \C
emission lines for reverberation-mapping AGNs in the literature. We
have measured them ourselves using the UV spectra from the IUE or HST
archive. The spectra of most objects are available from $IUE$, and a
few from $HST$. We measured the \Mg emission lines for 27 objects
and \C  emission lines for 33 objects.

Data reduction consists of several steps. First, all AGN spectra have
been extinction-corrected following a method given by Cardelli et
al. (1989) and shifted to the rest frame using the IRAF
software. Secondly iron emission features in each spectrum were
subtracted and the continuum is fitted with a power-law.  Thirdly the
FWHM and the flux of \Mg and \C  emission lines were measured.

\subsection{Iron emission subtraction and continuum fitting}

Without subraction of iron emission in UV/optical spectra of AGNs, the
accuracy of measurements of emission lines as well as the continua
(Boroson \& Green 1992; Corbin \& Boroson 1996; Vestergaard \& Wilkes
2001) would be limited, particularly for \Mg profile because it is
often blended by relative strong  Fe{\sc ii}  around 2800$\AA$.

Assuming the relative strength of iron lines (i.e. a template) are the
same for all AGNs, Boroson \& Green (1992) have developed a method to
fit and remove iron lines in optical spectra (in the rest frame) of
quasars.  Vestergaard \& Wilkes (2001) developed an iron template in
the UV band from $1250\AA$ to $3090\AA$ (in the rest frame) using HST
spectra of a narrow line Seyfert galaxy, I~Zwicky~1, which has rich
iron emission lines. They found that the template works well for
eliminating the iron emissions from spectra of a few quasars. We
followed the similar procedures as Vestergaard \& Wilkes (2001) to
subtract the iron emission using their iron template, and
simultaneously fit a power law to several continuum windows from
$1000\,\AA$ to $3100\,\AA$ without obvious emission lines or $
{Fe}{ii}$ emissions (see Table 2 from Kuraszkiewicz et al. 2002). We
finally determined the flux at $3000\AA$ or $1350\AA$ from
iron-subtracted spectrum.

\subsection{\C  and \Mg emission line measurements}

Now we try to obtain the emission line properties from the
clean-spectrum that both the iron emissions and the fitted continuum
were subtracted.

We normally use only one Gaussian component to fit the \C /\Mg
line profile and obtain FWHM, if there is no an obvious narrow
component. One component usually can provide a good description of
line profile, especially for \Mg. However, if an absorption line
is blended with \C  or \Mg lines, such as in NGC~4151,
PG~1411+442, another Gaussian absorption component is added in
fitting.  We ignore the asymmetry of \C  profiles often seen in
subtracted spectra, which may be induced by a relative broad
emission line such as He{\sc ii}$\lambda\,1640$ and
O{\sc {iii}]}$\lambda\,1663$ in the red-wing or an absorption line
in the blue-wing, as we believe that their net effect in fitting
result is small.  We measure the line width from each spectrum,
and then take the average for the FWHM of \Mg and \C  emission
lines (see Columns 4 and 7 in Table 1). We adopted an uncertainty of
20 percent if only one spectrum is available for an object. We did
not measure and take the line dispersion (the second moment of the
profile) as an alternative to the FWHM, which was suggested by
Peterson et al. (2004) and Fromerth \& Melia (2000), mainly because
the line dispersion agrees well with the FWHM in the limit of a
Gaussian line profile. As mentioned above, in most cases of our
measurements one Gaussian component can provide a satisfactory fit
to the line profile. Fluxes obtained by direct integration over the
observed emission line profile of the clean-spectrum were adopted.

In Table 1, we list all data used or obtained in this paper. The
columns are object name (1), $R_{BLR}$ (2) and the black hole mass (3)
from Peterson et al. (2004), the FWHM (4) and luminosity (5) of \Mg,
the luminosity of continuum at $3000\AA$ (6), the FWHM (7) and
luminosity (8) of \C , the continuum luminosity at $3000\AA$
(9). Data from column (4) to column (9) are our measurements (see the
following sections), and the number of the good spectra used is given
in columns (10) and (11) with marks indicating which archive of the
spectra.

\section{New empirical relations for black hole mass estimations}
We first try to establish the relations between the BLR size and the
luminosity of \Mg, \C  emission line, using data of the
reverberation mapping AGNs obtained above. These relations can be used
to estimate the BLR size of high-redshift AGNs from measurements of
the luminosity of \Mg, \C  emission line, as we will do in the next
section. Together with the BLR velocity obtained from the line-width,
the black hole masses of high-redshift AGNs can be estimated. In this
section, we will also use the known masses of black holes of the
reverberation mapping AGNs to ``calibrate'' the empirical relations.

\subsection{Relationships between the BLR size and \Mg, \C
emission line luminosity of the reverberation AGNs}

We take the value of the BLR size of the reverberation mapping sample
from Peterson et al. (2004) and the UV line luminosities measured
above to investigate their relations.
\begin{figure}
\begin{center}
\hspace{3mm}  \psfig{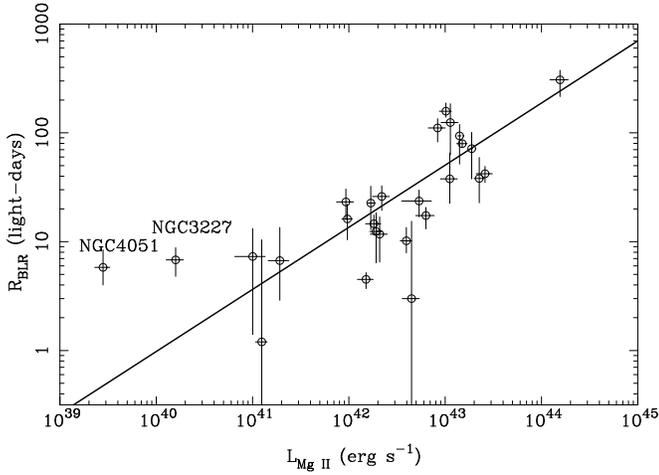}
  \caption{The BLR size, $R_{BLR}$, versus the luminosity of \Mg
    emission line ($\lambda\,2798\,\AA$), $L_{Mg{II}}$, of 27 AGNs in the
    reverberation mapping sample for which the \Mg emission line
    have been measured in Sect.2. The correlation coefficient between
    the two parameters is 0.72. The solid line shows an OLS bisector
    fit to data (Eq.~\ref{e1}), and has a slope of 0.57.}
  \label{Fig.1}
\end{center}
\end{figure}
\begin{figure}
\begin{center}
\hspace{3mm}  \psfig{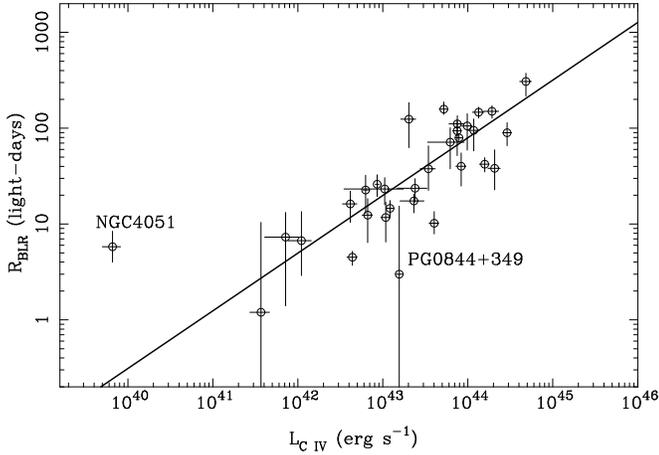}
  \caption{The same as Fig.\ref{Fig.1}, but for the \C  emission line
   ($\lambda\,1549\,\AA$) of 33 AGNs. The correlation coefficient is
   0.76. The solid line (Eq.~\ref{e2}) has a slope of 0.60.}
  \label{Fig.2}
\end{center}
\end{figure}
As shown in Fig.~\ref{Fig.1} for the \Mg line luminosities of 27
AGNs and Fig.~\ref{Fig.2} for the \C  line luminosities of 33 AGNs,
the BLR size and the line luminosities are closely related, with the
correlation coefficients of 0.72 and 0.76, respectively. The best
fitted lines with the OLS bisector method (Isobe et al. 1990) are
\begin{equation}
  \log\frac{R_{BLR}}{\rm   lt.days} = (1.13\pm0.13)+
  (0.57\pm0.12)\log(\frac{L_{Mg{II}}}{10^{42}{\rm erg\,s^{-1}}}),
\label{e1}
\end{equation}
\begin{equation}
  \log\frac{R_{BLR}}{\rm   lt.days} = (0.69\pm0.28)+
  (0.60\pm0.16)\log(\frac{L_{C{IV} }}{10^{42}{\rm erg\,s^{-1}}}).
\label{e2}
\end{equation}
Given the uncertainties,  the slopes are slightly larger than but
still consistent with the value 0.5 expected from a simple
photo-ionization model for the BLR.

We also investigated the relations between the $R_{BLR}$ and the
continuum luminosity $\lambda\,L_{3000\AA}$ or $\lambda\,L_{1350\AA}$
using our measurements in Table 1, and obtained the following
relations,
\begin{equation}
  \log\frac{R_{BLR}}{\rm   lt.days} = (1.27\pm0.10)+
  (0.58\pm0.10)\log(\frac{L_{3000\AA}}{10^{44}{\rm erg\,s^{-1}}}),
\label{e3}
\end{equation}

\begin{equation}
  \log\frac{R_{BLR}}{\rm   lt.days} = (1.15\pm0.14)+
  (0.56\pm0.12)\log(\frac{L_{1350\AA}}{10^{44}{\rm erg\,s^{-1}}}).
\label{e4}
\end{equation}
The correlation coefficients are 0.72 and 0.75, respectively. For the
relationship between $\rm R_{BLR}$ and $\rm L_{3000\AA}$, the slope
of $0.58\,\pm\,0.10$ we obtained is consistent with $0.47\,\pm\,0.05$
derived by McLure \& Jarvis (2002) from
34 reverberation AGNs (Kaspi et al. 2000).
The relationship between $\rm R_{BLR}$ and $\rm L_{1350\AA}$ was
obtained recently by Kaspi et al. (2005) for the first time from
measurements, with a slope of $0.56\,\pm\,0.05$ from the BCES method
and $0.50\,\pm\,0.04$ from the $fitexy$ method. Our slope of
$0.56\,\pm\,0.12$ is well consistent with their value.

There is a well-known relationship between the equivalent width of \C
broad emission line and the 1350$\AA$ continuum luminosity of AGNs (Baldwin
1977). This so-called Baldwin-effect can be converted to $L_{\C} \propto
L_{1350\AA}^{\alpha}$, with a typical value of $\alpha\sim0.4$ for an
individual AGN but $\alpha\sim0.83\pm0.04$ for a sample of AGNs (see
Peterson 1997 and references therein). Using our measurements of
luminosities of $L_{1350\AA}$ and $L_{\C}$ for 33 reverberation mapping
AGNs, we got $\alpha = 0.94 \pm 0.06$. Given $R_{BLR} \propto
L_{\C}^{0.60\pm0.16}$ as in Eq. (2), we can easily derive $R_{BLR} \propto
L_{1350\AA}^{0.56}$, which is exactly shown in Eq.(4).

Exclusion of four radio loud objects ($R\,>\,10$) from the 35 AGNs in
the reverberation sample in Kaspi et al. (2005), i.e. 3C~120, 3C390.3,
IC~4329A and PG~1226+023, only causes a small change for the slope of
$R\,-\,L$ relations derived by Kaspi et al. (2005) and by us, because
the sample is dominated by 31 radio quiet objects (Nelson 2000).

\subsection{Black hole mass estimate and calibration}

The mass of a black hole can be determined by the $R_{\rm BLR}$ and
the velocity at this radius. The best is to use the time delay of
variation of the  \Hb  line to determine the $R_{\rm BLR}$, and also
use the  \Hb  line-width for the velocity estimate. In this paper, to
estimate the black hole mass, we will use the $R_{\rm BLR}$ calculated
from luminosities of \Mg and \C  lines using Eq. (1) and Eq. (2),
which are ``calibrated'' from the measurements of  \Hb  line, and also
use the FWHM of \Mg and \C  lines for the velocity estimates. Note
that the time lags of flux variations of different emission lines
related to that of the UV/optical continuum are different, as shown
from the measurements of the reverberation AGNs (Peterson et al. 2000,
2005; Onken \& Peterson 2002; Korista et al. 1995). For example, the
time lag for the \C  line variation of NGC 5548 is approximately half
of that for the  \Hb  line, which indicates that the different
emission lines are probably emitted at the different distances from
the center of an AGN.  Therefore we need to ``calibrate'' relations
for black hole mass estimation when we have $R_{BLR}$ and $V$ from
different emission lines.
\begin{figure}
\begin{center}
\hspace{3mm}  \psfig{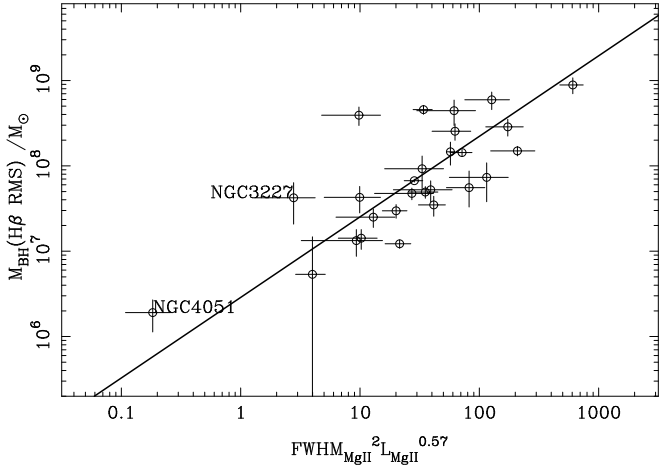}
  \caption{The black hole masses of 27 reverberation AGNs derived from
    the  \Hb  emission line from Peterson et al. (2004) versus the
    ``mode'' black hole masses from the FWHM and luminosity of \Mg
    emission line.  The slope of the line is $0.94\pm0.15$ from an OLS
    bisector fit, and the correlation coefficient is 0.76. Here the
    FWHM$_{Mg{II}}$ is in units of 1000~km~s$^{-1}$, and
    $L_{Mg{II}}$ in units of 10$^{42}$~erg~s$^{-1}$.}
  \label{Fig.3}
\end{center}
\end{figure}
\begin{figure}
\begin{center}
\hspace{3mm}\psfig{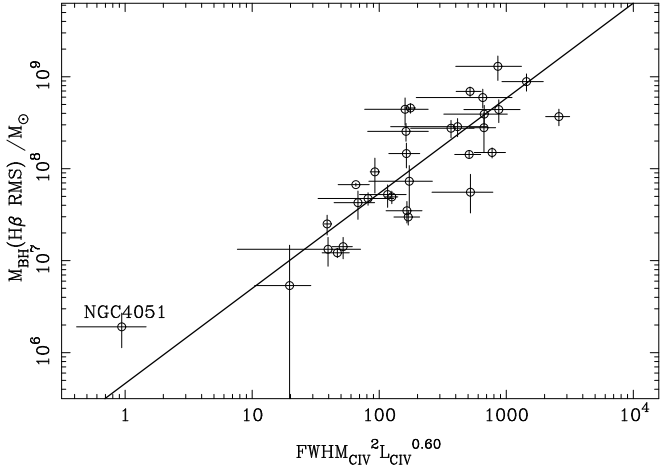}
  \caption{Same as Fig.~\ref{Fig.3} but for \C  emission lines of 33
      reverberation AGNs.  The slope of the line is $1.0\pm0.16$ from
      an OLS bisector fit, and the correlation coefficient is 0.86. }
  \label{Fig.4}
\end{center}
\end{figure}

Similar to Vestergaard (2002), we can estimate $R_{\rm BLR}$ from the line
luminosity of either $L_{\Mg}$ or $L_{\C}$ (see Eq. (1) -- Eq. (2)), and the
velocity from the FWHM of \Mg or \C . Finally we estimate the black hole
masses from the $R_{\rm BLR}$ and FWHM value. Using the known black hole
masses of the reverberation AGNs in Peterson et al. (2004), we obtained the
``calibrated'' relations (shown in Fig.~\ref{Fig.3} and Fig.~\ref{Fig.4})
for black hole masses as the following:
\begin{equation}
  M_{BH}(Mg{II}) = 2.9 \times 10^{6} \left
  (\frac{L_{Mg{II}}}{\rm 10 ^{42} erg s^{-1}}\right )^{0.57\pm0.12}\left
  [\frac{\rm FWHM_{Mg{II}}}{\rm 1000 km s^{-1}}\right ]^{2} M_{\sun},
\label{e5}
\end{equation}

\begin{equation}
  M_{BH}(C{IV}) = 4.6 \times 10^{5} \left
  (\frac{L_{C{IV}}}{\rm 10 ^{42} erg s^{-1}}\right )^{0.60\pm0.16}\left
  [\frac{\rm FWHM_{C{IV}}}{\rm 1000 km s^{-1}}\right ]^{2} M_{\sun}.
\label{e6}
\end{equation}
%

We also tried to estimate the black hole masses using the $R_{\rm BLR}$
estimated from continuum luminosities at $3000\,\AA$ and $1350\,\AA$ and the
FWHM of \Mg and \C  lines. We got the following relations:
%
\begin{equation}
M_{BH}(3000\AA) = 3.4\times10^{6}\,\left
(\frac{\lambda\,L_{3000\,\AA}}{\rm 10 ^{44}\,erg\,s^{-1}}\right
)^{0.58\pm0.10}\left [\frac{\rm FWHM_{{Mg}{II}}}{\rm 1000\,km\,s^{-1}}\right
]^{2}M_{\sun},
\label{e7}
\end{equation}
\begin{equation}
M_{BH}(1350\AA) = 1.3\times10^{6}\,\left
  (\frac{\lambda\,L_{1350\,\AA}}{\rm 10 ^{44}\,erg\,s^{-1}}\right
  )^{0.56\pm0.12}\left [\frac{\rm FWHM_{{C}{IV}}}{\rm 1000\,km\,s^{-1}}\right
  ]^{2}M_{\sun}.
\label{e8}
\end{equation}
Note Eq.(7) is  similar to the Eq.(7) in McLure \& Jarvis (2002) who
obtained $M_{BH}(3000\AA) \propto L_{3000\,\AA}^{0.47} {\rm FWHM}_{\eMg}^2$.
Our slope obtained with 27 reverberation mapping AGNs is slightly larger
than that they obtained with 22 objects.
%
Eq.(6) and Eq.(8) are very useful to calculate black hole masses of
high redshift AGNs. Vestergaard (2002) has tried to do it using the
$R_{\rm BLR}$ estimated from the continuum luminosity at $1350\,\AA$
and the FWHM of \C  line, adopting $R_{\rm BLR} \propto (\lambda
L_{1350\AA})^{0.7}$.  The black hole mass we obtained using Eq.(8) is
slightly lower than that Vestergaard (2002) derived, but consistent
with each other within one order of magnitude.

\section{Black hole mass estimation for several AGN samples}

The relations obtained in last section for UV lines can be used to estimate
black hole masses of high redshift AGNs. For radio loud objects, as
argued in Wu et al. (2004), jets could affect the estimate of $R_{\rm BLR}$
from the continuum luminosities. Therefore, our relations in Eqs. (1), (2),
(5) and (6) based on the line luminosity for $R_{\rm BLR}$ should give
better estimates of black hole masses for radio loud AGNs than those based on the
continuum luminosities.

We apply the relations obtained above to estimate the black hole masses of
AGNs in two samples. The first one is the Large Bright Quasar Survey (LBQS)
sample (Forster et al. 2001;
Hewett et al. 1995; Hewett et al. 2001). The second one is a composite
radio-loud AGN sample (Cao \& Jiang 1999; Celotti et al. 1997; Barthel et
al. 1990; Constantin et al. 2002).

\subsection{Data for LBQS sample}
Forster et al. (2001) measured the optical/UV continuum and emission line
properties of a homogeneous sample of 993 quasars in the LBQS, including the
FWHM and equivalent width of lines, the flux and slope of continuum. The
continuum fluxes have been Galactic extinction corrected. There are 504
objects with \Mg line measurements and 403 objects with \C  line
measurement. The black hole masses of these quasars can be estimated using
the relations presented in Section 3. The continuum luminosities were
calculated after K-correction according to the continuum slope. The line
luminosities were calculated from the equivalent width of the emission line.

We also obtained the radio-loudness for these quasars.
Radio-loudness is originally defined as the ratio between the radio
flux density at 5~GHz and optical flux density at B band
($4400\,\AA$) (Kellermann et al. 1989). We adopt $B_{J}$ magnitude
(Hewett et al. 1995) as an approximation of B band magnitude, which
only yields an increase of 0.03 in $\log R$ on average as Hewett et
al. (2001) got the mean $B-B_{J}\simeq0.07$. The optical luminosity
at 4400\,{\AA} at the rest frame of a source were calculated by
assuming an optical continuum slope of 0.3
($f_{\nu}\,\propto\,\nu\,^{-\,\alpha}$, Schmidt \& Green 1983). For
radio data, we made the cross-identifications of quasars to sources
in the catalog of the NRAO VLA Sky Survey (Condon et al. 1998) and
found radio counterparts for 84 quasars with \Mg line measurements
and 60 quasars with \C  line measurements. The flux density
observed at 1.4~GHz was then scaled to flux at 5~GHz at the rest
frame of a source by roughly assuming a radio continuum slope of
0.5. So-obtained optical and radio data were used to calculate the
radio loudness.

\subsection{Data for a composited sample of radio loud AGNs}
Celotti et al. (1997) collected a radio-loud sample to investigate the
relation between the broad-line emission and the central engine of
AGNs. Similarly, Cao \& Jiang (1999) assembled a sample of radio-loud
quasars and BL Lac objects from $\rm 1-JY$ S4 and S5 radio source catalogs
to investigate the relations between the broad emission line luminosity and
jet power. Barthel et al. (1990) presented optical spectra of 67 radio-loud
quasars ($1.5\,<\,z_{em}\,<\,3.8$) and measured the line properties,
including \Mg and \C  lines. Recently Constantin et al. (2002) observed
the \C  emission line properties of 34 quasars at high-redshift
($z\,>\,4$).

Using all these data, we composite a sample of radio-loud AGNs. We
thus got 126 radio-loud AGNs with the \Mg line luminosity and 164
AGNs with the \C  line luminosity. While, the FWHM data of the
\Mg line are only available for 86 objects, and \C  line for 92
objects.  The radio-loudness of these AGNs were estimated using the
data of 5 GHz flux density and absolute magnitude $M_{B}$ available in
the catalog of V\'{e}ron-Cetty \& V\'{e}ron (2003). Optical flux
densities were corrected for extinction using the $\rm A_{B}$ value
available from NED, and also K-corrected assuming an optical spectral
index of 0.3. Radio flux density is also K-corrected assuming a slope
of 0.5 if no radio spectral index is available in literature.

\subsection{Black hole masses from the $\RMg$ and $\RcMg$ relations}

We got 210 AGNs with the \Mg measurements from two samples.  But
$\rm M_{BH}$(\Mg) and $\rm M_{BH}(3000\AA)$ can be estimated using the
relations in Eqs. (1),(3),(5) and (7) for 170 objects of them having
the available FWHM data of the \Mg line.

\begin{figure}
\begin{center}
\hspace{3mm}\psfig{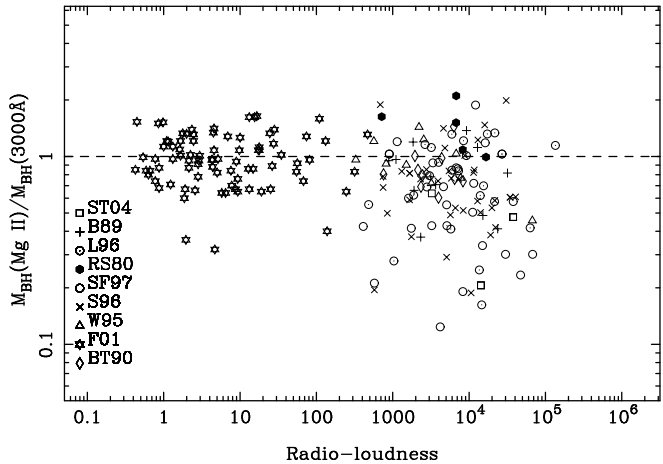}
  \caption{The ratio of black hole masses estimated from the
    luminosities of \Mg and continuum versus radio-loudness for 210
    AGNs.  Different symbols indicate objects from different
    references: Sbarufatti et al. (2004, ST04); Baldwin Wampler \&
    Gaskell (1989, B89); Lawrence et al. (1996, L96); Richstone \&
    Schmidt (1980, RS80); Scarpa \& Falomo (1997, SF97); Stickel \&
    Kuehr (1996, S96); Wills et al. (1995, W95); Forster et al. (2001,
    F01); Barthel, Tytler \& Thomson (1990, BT90). The dashed line
    indicates the identical black hole masses. 
    $z>1.0$.  }
  \label{Fig.5}
 \end{center}
\end{figure}

\begin{figure}
\begin{center}
\hspace{3mm}\psfig{file=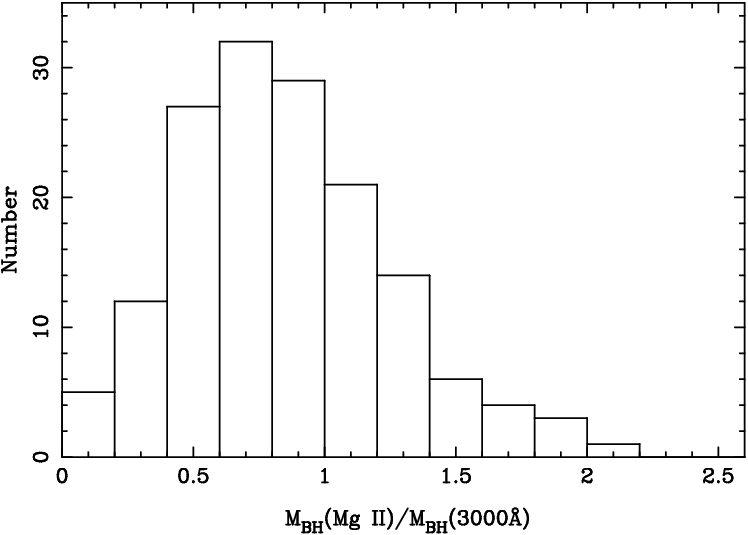,width=87mm,angle=270}
  \caption{Histogram of $\rm M_{\eMg}/M_{BH}(3000\AA)$ for the 154
    radio-loud AGNs ($R\,>\,10$) shown in Fig.~\ref{Fig.5}. About
    69\% of the objects have the mass ratio less than 1. The average
    mass ratio is 0.80 with $\sigma\,=\,0.1$.}
  \label{Fig.6}
 \end{center}
\end{figure}

Now we investigate the correlation between the mass ratio, $\rm
M_{BH}$(\Mg)/$\rm M_{BH}(3000\AA)$, and the radio-loudness for the 210
AGNs. For this purpose, the FWHM data are not necessary because $\rm
M_{BH}$(\Mg)/$\rm M_{BH}(3000\AA)$ calculated from Eqs. (5) and (7)
only depends on the ratio of $L_{\eMg}$ and $L_{3000\AA}$. As shown in
Fig.~\ref{Fig.5}, when the radio-loudness is small, black hole masses
estimated from luminosities of line and continuum are statistically
identical. When the radio-loudness increases, the mass ratio tends to
be more scattered and systematically lower than 1.0. Specifically, the
mass ratio is less than 0.4 for five BL Lac objects (S5 1803+78, 4C
56.27, PKS 0537-441, PKS 1144-379, PKS 2029+121).  Fig.~\ref{Fig.6}
shows the histogram of the $\rm M_{BH}$(\Mg)/$\rm M_{BH}(3000\AA)$ of
the 154 radio-loud AGNs. The mass ratio on average is 0.80 with a
deviation of $\sigma=0.1$. This result from these two samples confirms
the suggestion in Wu et al. (2004) that for radio-loud objects,
especially extremely radio-loud AGNs, black hole masses should be
estimated from the $R_{BLR}$ - line luminosity relation.
\begin{figure}
\begin{center}
\hspace{3mm} \psfig{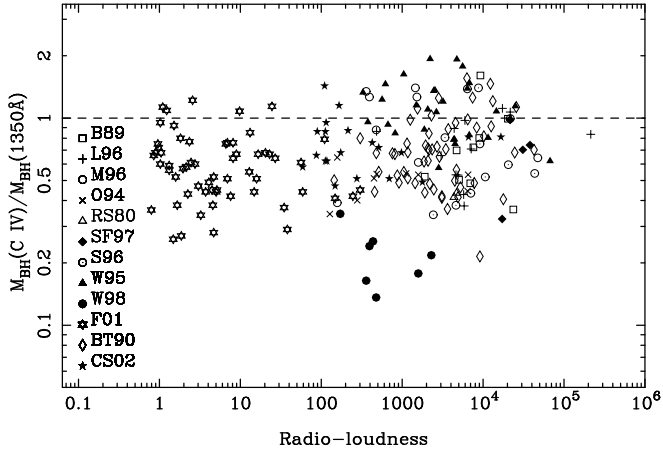}
  \caption{Same as Fig.~\ref{Fig.5}, but for 224 AGNs with measurement
    of \C  line. Different symbols indicate objects from different
    references: Baldwin Wampler \& Gaskell (1989, B89); Lawrence et
    al. (1996, L96); Marziani et al. (1996, M96); Osmer, Porter \&
    Green (1994, O94); Richstone \& Schmidt (1980, RS80); Scarpa \&
    Falomo (1997, SF97); Stickel \& Kuehr (1996, S96); Wills et
    al. (1995, W95); Wang et al. (1998, W98); Forster et al. (2001,
    F01); Barthel, Tytler \& Thomson (1990, BT90); Constantin et
    al. (2002, CS02). The ratio of the black hole masses were
    estimated from the $\RC$ and $\RcC$ relations. }
  \label{Fig.7}
\end{center}
\end{figure}

\begin{figure}
\begin{center}
\hspace{3mm}
  \psfig{file=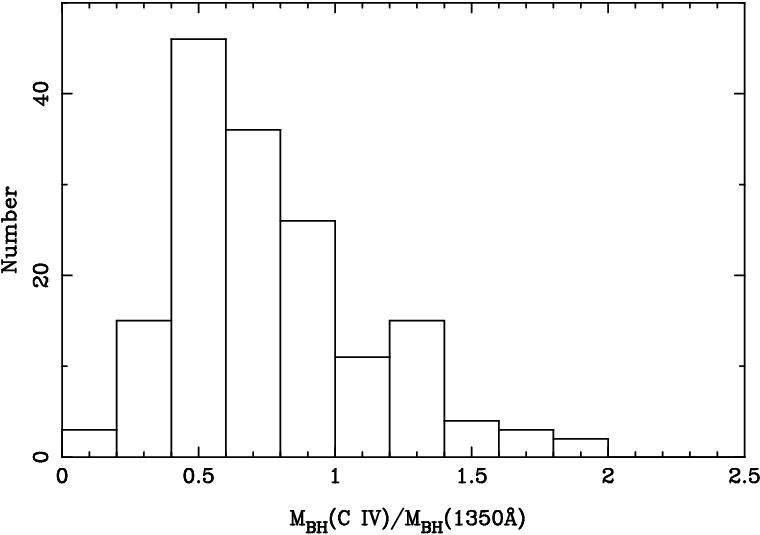,width=87mm,angle=270}
  \caption{Histogram of $\rm M_{BH}(\C)/M_{1350\,\AA}$ for the 181
    radio-loud AGNs ($R\,>\,10$) shown in Fig.~\ref{Fig.7}. About
    70\% of the objects have the mass ratio less than 1. The average
    mass ratio is 0.73 with $\sigma\,=\,0.2$.}
  \label{Fig.8}
\end{center}
\end{figure}

\subsection{Black hole masses from the $\RC$ and $\RcC$ relations}

We got 224 AGNs with the \C measurement from two samples.  Similarly
to the \Mg sample, we can estimate the black hole masses for 152
objects which have the FWHM data of \C line. To check the dependence
of the mass ratio, $\rm M_{BH}$(\C)/$\rm M_{BH}(1350\AA)$, on the
radio-loudness, we can use all 224 AGNs, as shown in
Fig.~\ref{Fig.7}. Fig.~\ref{Fig.8} shows the histogram of $\rm
M_{BH}$(\C)/$\rm M_{BH}(1350\AA)$ for the radio-loud 181 AGNs ($R>
10$). Clearly the black hole masses estimated from the BLR size
derived from the $\RcC$ relation are probably overestimated as about
70\% of the objects in the sample have the mass ratio less than 1.
The average mass ratio is 0.73 with $ \sigma=0.2$. Therefore, it is
confirmed again that the continuum luminosity is not a good
indication of ionizing luminosity due to the possible jet
contribution.

\section{Discussions and Conclusions}

Black hole masses of AGNs can be estimated either from the relations
between the BLR size and UV emission line luminosity or the relations
between the BLR size and UV continuum luminosity. We have obtained the
new empirical relations between the BLR size and the \Mg, \C
emission line luminosity from the reverberation-mapping AGNs (see
Eq.~(\ref{e1}), Eq.~(\ref{e2})). This enables us to estimate the black
hole masses of high-redshift ($z>0.8$) AGNs, using the luminosity and
the FWHM of these UV emission lines (see Eq.~(\ref{e5}),
Eq.~(\ref{e6})).

The so-obtained black hole masses were compared to those obtained
using the relations between the BLR size and UV continuum luminosity
(see Eq.~(\ref{e7}), Eq.~(\ref{e8})). For radio-loud AGNs, jets
contribute substantially to the continuum luminosity, therefore the
black hole masses estimated from the BLR size - continuum luminosity
relations ( Eq.~(\ref{e3}), Eq.~(\ref{e4})) are probably
overestimated. The relation of $R_{\rm BLR}$ - $L_{\eC}$ or that of
$R_{\rm BLR}$ - $ L_{\eMg}$ should be a better choice.

For AGNs with the measurements of \Mg lines, we have shown that $\rm
M_{BH}$(\Mg) is very close to $\rm M_{BH}(3000\AA)$ when the
radio-loudness is small. The black hole masses of radio loud AGNs
estimated from the continuum luminosity are systematical larger than
those from the line luminosity, confirming the suggestion by Wu et
al. (2004) for $\rm M_{BH}($\Hb$)\,/\,M_{BH}(5100\AA)$.

For AGNs with the measurements of \C  lines, our estimates for black hole
masses of LBQS
sample are slightly smaller than but still consistent with those estimated
using the relations given by Vestergaard (2002), with a slope of 0.92 and a
correlation coefficient of about 0.9. The distribution of the mass ratio
does not show a strong correlation with the radio-loudness. We noticed that
Baskin \& Laor
(2005) mentioned  for the high ionization emission line like \C
the gravitational effects on the emission line shift and profile often can
not be seen,  while for low ionization
line like \Mg and  \Hb  this effects often can be seen. If this is true, it
implies that the physics of
high-ionization \C  emission lines may be different from the low ionization
lines. Therefore, we should be cautious when estimating the black hole mass
with \C  emission line properties.

\begin{figure}
\begin{center}
\hspace{3mm}\psfig{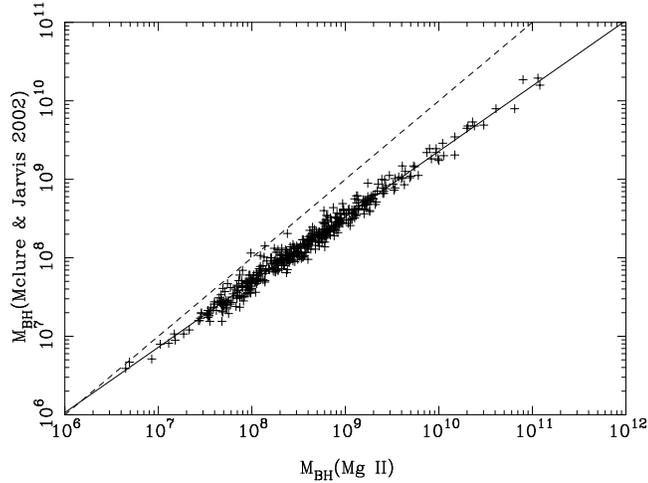}
  \caption {Comparison of black hole masses estimated from luminosity
   of \Mg (our Eq.(5)) and continuum at $3000\AA$ (Mclure \& Jarvis
   2002) using LBQS sample. 476 objects were used which $\rm
   1000\,km\,s^{-1} < FWHM_{\eMg} < 10000\,km\,s^{-1}$ for avoiding
   measurement errors or narrow line contamination. The linear fit
   gives a slope of 0.83 (solid line) with a coefficients of 0.99. The
   dashed line indicates the equal masses from two approaches. }
  \label{Fig.9}
\end{center}
\end{figure}
\begin{figure}
   \begin{center}
\hspace{3mm}  \psfig{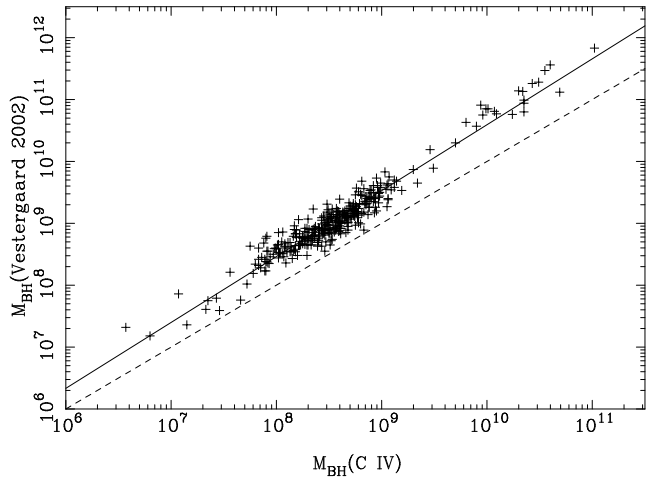}
  \caption{ Same as Fig~\ref{Fig.9} but for black hole masses estimated from
   luminosity of \C  (our Eq.(6)) and continuum at $1350\AA$ (Vestergaard 2002)
   with 341 objects plotted. The slope of the fit (solid line) is 1.06 and
   the correlation coefficients is 0.96.  }
  \label{Fig.10}
\end{center}
\end{figure}

We have compared the black hole masses of LBQS quasars estimated using
Eq. (5) and Eq. (6) with those obtained by McLure \& Jarvis (2002) and
Vestergaard (2002). Fig. 9 and Fig. 10 show the comparisons of these
results. Our estimates are systematically larger than those in McLure
\& Jarvis (2002) and smaller than those in Vestergaard (2001). This
mainly comes from the different slopes of the R-L relations used in
these studies.

It is important to understand the uncertainties of data used to obtain
the relations presented in this paper. These uncertainties come from
the errors of the measured BLR size, the variation of emission line
flux and the line FWHM, and the different inclination of the
BLR. These factors have been already discussed by Wu \& Liu
(2004). The BLR inclination will not affect the results in this work
since the black hole mass ratio $\rm M_{BH}$(\Mg)/$M_{BH}(3000\AA)$ and
$\rm M_{BH}$(\C)/$M_{BH}$(1350\AA) has nothing to do with
inclination. The plots for the BLR size and luminosity seem to have a
larger scatter for lower luminosity objects (see Fig. (1) and Fig.
(2)), which can also be seen in previous studies (Kaspi et al. 2000;
Wu et al. 2004). More data of the low luminosity AGNs, especially with
future reverberation mapping measurements, should be collected to
improve the suggested relations.

Recently, more and more high redshift or high luminous quasars or
radio-loud ones have been found.  Because recent studies have used
such $R\,-\,L$ relations to obtain black hole masses of high redshift
quasars with $4\,\leq\,z\,\leq\,6$ (Vestergaard 2004), including the
most distant quasar $\rm SDSS J114816.64\,+\,525150.3$ (z\,=\,6.4)
(Willott, McLure \& Jarvis 2003; Barth et al. 2003). However, the
relations for $R_{\rm BLR}$ in Eq.(1,2,3,4) were obtained from the
AGNs in reverberation sample, most of which are not very luminous and
radio-quiet AGNs. We need to be caution if the black hole masses of
these high redshift quasars are overestimated, especially when they
are radio-loud objects. Indeed, a few high-z quasars have been already
identified as radio-loud quasars (Fan et al. 2001; Momjian, Petric \&
Carilli 2004) therefore, further checks on the validity of suggested
R-L relations for AGNs at high redshift or with large luminosity ones
or large radio-loudness are still needed.

\section*{acknowledgements}
We thank Dr. Marianne Vestergaard for kindly providing us her iron
template data in the UV band, and Dr. Michael Corbin for useful
suggestions on the template and sending us his iron template spectra
around $3000\AA$.
KMZ is grateful to Xiaohui Sun, Jing Wang, Yong Zhang and Yougang Wang
for their numerous helps on fitting the spectra.
XBW is supported by the National Natural Science Foundation of China
(No. 10473001 \& No. 10525313) and the Research Fund for the
Doctoral program of Higher Education (No. 20050001026).
JLH is supported by the National Natural Science Foundation of China
(10025313) and the National Key Basic Research Science Foundation of
China (G19990754) as well as the partner group of MPIfR at the
National Astronomical Observatories of China.
This work is based on the INES data from $IUE$ satellite and the data from
$HST$ satellite partly.
This research has made use of the NASA/IPAC Extragalactic Database
(NED) which is operated by the Jet Propulsion Laboratory, California
Institute of Technology, under contract with the National Aeronautics
and Space Administration.






\clearpage
\begin {center}
 \begin{table}
\begin {small}
 \tabcolsep 0.8pt \renewcommand\arraystretch{1.5}
 \centering
 \setlength{\unitlength}{-10.1pt}

 \caption{\label{para}Parameters of \Mg, \C  emission lines for 35
AGNs of reverberation-mapping AGN sample.  }

 \begin{tabular}{llllllllllcc}
 \hline \hline \noalign{\smallskip}

 $ $&$\rm R_{BLR}$&$\rm M_{\rm BH}$&$\rm FWHM_{\Mg}$&$L_{\Mg}$&$L_{3000\AA}$ &$
\rm FWHM_{\C}$&$L_{\C}$&$L_{1350\AA}$&Num&Num\\
 Object&(lt-days)&$10^{6}M_{\sun}$&$\rm
(km\,s^{-1})$&$(\rm 10^{42}\,erg\,s^{-1})$&$(\rm 10^{44}\,erg\,s^{-1})$&$\rm
(km\,s^{-1})$&$\rm (10^{42}\,erg\,s^{-1})$&$(\rm
10^{44}\,erg\,s^{-1})$&\Mg&\C \\
(1)&(2)&(3)&(4)&(5)&(6)&(7)&(8)&(9)&(10)&(11)\\

 \noalign{\smallskip} \hline \noalign{\smallskip}

3C120 & 38.1$^{+21.3}_{-15.3}$ & 55.5 $^{+31.4}_{-22.5}$ &
3740$\pm$538& 22.57$\pm$2.64& 10.44$\pm$2.63& 4652$\pm$937&
206.81$\pm$34.22& 34.30$\pm$10.05&13 &34 \\ 3C390.3 &
23.6$^{+6.2}_{-6.7}$ & 287$\pm$64 & 8174$\pm$622& 5.36$\pm$1.83&
0.99$\pm$0.48& 7884$\pm$1870& 23.91$\pm$9.14& 2.41$\pm$1.30&6 &92 \\
Akn120 & 42.1$^{+7.1}_{-7.1}$ & 150$\pm$19 & 5727$\pm$842&
25.93$\pm$4.94& 15.45$\pm$3.10& 6134$\pm$603& 157.83$\pm$20.76&
46.75$\pm$10.21&20 &37 \\ F9 & 17.4$^{+3.2}_{-4.3}$ & 255$\pm$56 &
4682$\pm$547& 6.31$\pm$1.38& 2.94$\pm$0.93& 4981$\pm$782&
23.23$\pm$7.24& 3.67$\pm$2.27&54 &106 \\ MRK79 & 11.7$^{+5.2}_{-5.2}$
& 52.4$\pm$14.4 & 5072$\pm$1014& 2.09$\pm$0.42& 0.80$\pm$0.24&
5301$\pm$858& 10.80$\pm$1.36& 1.92$\pm$0.74&1 &4 \\ MRK110 &
26.0$^{+6.6}_{-6.6}$ & 25.1$\pm$6.1 & 2878$\pm$575& 2.19$\pm$0.44&
0.55$\pm$0.16& 3283$\pm$ 33& 8.59$\pm$0.45& 0.67$\pm$0.01&2 &2 \\
MRK279 & 12.4$^{+6.0}_{-6.0}$ & 34.9$\pm$9.2 & 5346$\pm$465&
1.92$\pm$0.27& 1.04$\pm$0.47& 7306$\pm$853& 6.63$\pm$0.91&
1.06$\pm$0.44&6 &16 \\ MRK335 & 14.6$^{+3.0}_{-3.0}$ & 14.2$\pm$3.7 &
2700$\pm$346& 1.82$\pm$0.32& 1.77$\pm$0.23& 3419$\pm$202&
12.17$\pm$1.27& 2.64$\pm$0.41&17 &22 \\ MRK509 & 79.6$^{+6.1}_{-5.4}$
& 143$\pm$12 & 3921$\pm$272& 15.03$\pm$1.87& 7.36$\pm$1.37&
6138$\pm$537& 78.66$\pm$7.43& 13.39$\pm$2.92&33 &50 \\ MRK590 &
23.2$^{+7.3}_{-7.3}$ & 47.5$\pm$7.4 & 5324$\pm$1064& 0.93$\pm$0.19&
0.15$\pm$0.05& 4477$\pm$441& 10.50$\pm$7.01& 0.85$\pm$0.71&1 &2 \\
MRK817 & 22.6$^{+9.7}_{-9.7}$ & 49.4$\pm$7.7 & 5112$\pm$601&
1.69$\pm$0.12& 1.07$\pm$0.04& 6481$\pm$236& 6.27$\pm$0.44&
1.28$\pm$0.13&2 &3 \\ NGC3227 & 6.8$^{+2.0}_{-2.0}$ & 42.2$\pm$21.4 &
5406$\pm$1081& 0.02$\pm$0.00&
0.01$\pm$0.00&.\,\,\,\,.\,\,\,\,.\,\,\,\,.\,\,\,\,.&.\,\,\,\,.\,\,\,\,.\,\,\,\,.\,\,\,\,.&.\,\,\,\,.\,\,\,\,.\,\,\,\,.\,\,\,\,.&1
&0 \\ NGC3516 & 6.7$^{+6.8}_{-3.8}$ & 42.7$\pm$14.6 & 5036$\pm$900&
0.19$\pm$0.05& 0.15$\pm$0.05& 8018$\pm$681& 1.10$\pm$0.33&
0.19$\pm$0.09&19 &76 \\ NGC3783 & 10.2$^{+3.3}_{-2.3}$ & 29.8$\pm$5.4
& 3031$\pm$235& 3.96$\pm$0.54& 2.03$\pm$0.50& 4320$\pm$329&
40.13$\pm$5.10& 8.35$\pm$2.23&97 &76 \\ NGC4051 & 5.8$^{+2.6}_{-1.8}$
& 1.91$\pm$0.78 & 2273$\pm$350& 0.003$\pm$0.001& 0.004$\pm$0.001&
4337$\pm$898& 0.01$\pm$0.002& 0.002$\pm$0.0004&8 &19 \\ NGC4151 &
7.3$^{+5.9}_{-5.9}$ & 13.3$\pm$4.6 & 5866$\pm$1342& 0.10$\pm$0.03&
0.06$\pm$0.04& 6939$\pm$1902& 0.72$\pm$0.31& 0.07$\pm$0.05&416 &454 \\
NGC4593 & 1.2$^{+9.2}_{-5.3}$ & 5.36$^{+9.37}_{-6.95}$ & 3613$\pm$363&
0.12$\pm$0.02& 0.08$\pm$0.02& 5983$\pm$939& 0.37$\pm$0.09&
0.05$\pm$0.02&13 &25 \\ NGC5548 & 16.2$^{+5.8}_{-5.8}$ & 67.1$\pm$2.6
& 5394$\pm$288& 0.97$\pm$0.12& 0.36$\pm$0.10& 5289$\pm$421&
4.15$\pm$0.80& 0.41$\pm$0.38&116 &179 \\ NGC7469 & 4.5$^{+0.7}_{-0.8}$
& 12.2$\pm$1.4 & 4123$\pm$284& 1.51$\pm$0.28& 0.81$\pm$0.20&
4402$\pm$352& 4.38$\pm$0.60& 1.06$\pm$0.17&11 &229 \\
PG0026+129&111.0$^{+24.1}_{-28.3}$ & 393$\pm$96 & 1714$\pm$342&
8.33$\pm$1.67& 19.27$\pm$5.78& 7131$\pm$1426& 75.22$\pm$15.04&
23.28$\pm$6.98&1 &1$^{a}$ \\ PG0052+251& 89.8$^{+24.5}_{-24.1}$ &
369$\pm$76
&.\,\,\,\,.\,\,\,\,.\,\,\,\,.\,\,\,\,.&.\,\,\,\,.\,\,\,\,.\,\,\,\,.\,\,\,\,.&.\,\,\,\,.\,\,\,\,.\,\,\,\,.\,\,\,\,.&
9380$\pm$690& 289.44$\pm$32.21& 29.74$\pm$2.25&0 &2 \\
PG0804+761&146.9$^{+18.8}_{-18.9}$ & 693$\pm$83
&.\,\,\,\,.\,\,\,\,.\,\,\,\,.\,\,\,\,.&.\,\,\,\,.\,\,\,\,.\,\,\,\,.\,\,\,\,.&.\,\,\,\,.\,\,\,\,.\,\,\,\,.\,\,\,\,.&
5276$\pm$342& 134.24$\pm$20.95& 29.28$\pm$4.62&0 &2 \\ PG0844+349&
3.0$^{+12.4}_{-10.0}$ & 92.4$\pm$38.1 & 3759$\pm$751& 4.48$\pm$0.90&
4.61$\pm$1.38& 4237$\pm$ 60& 15.56$\pm$1.42& 5.77$\pm$0.08&1$^{a}$&2
\\ PG0953+415&150.1$^{+21.6}_{-22.6}$ & 276$\pm$59
&.\,\,\,\,.\,\,\,\,.\,\,\,\,.\,\,\,\,.&.\,\,\,\,.\,\,\,\,.\,\,\,\,.\,\,\,\,.&.\,\,\,\,.\,\,\,\,.\,\,\,\,.\,\,\,\,.&
3987$\pm$797& 192.41$\pm$38.48& 45.61$\pm$13.68&0 &1$^{a}$ \\
PG1211+143& 93.8$^{+25.6}_{-42.1}$ & 146$\pm$44 & 3569$\pm$ 79&
14.10$\pm$0.58& 11.72$\pm$0.09& 3527$\pm$365& 74.81$\pm$8.33&
20.09$\pm$2.00&2$^{a}$&11 \\ PG1226+023&306.8$^{+68.5}_{-90.9}$ &
886$\pm$187 & 5874$\pm$292& 155.74$\pm$33.52& 115.31$\pm$31.46&
5998$\pm$794& 482.92$\pm$75.31& 206.81$\pm$41.58&3$^{a}$&207 \\
PG1229+204& 37.8$^{+27.6}_{-15.3}$ & 73.2$\pm$35.2 & 5414$\pm$1083&
11.15$\pm$2.23& 3.17$\pm$0.95& 4562$\pm$912& 34.44$\pm$6.89&
15.88$\pm$4.76&1 &1 \\ PG1307+085&105.6$^{+36.0}_{-46.6}$ &
440$\pm$123
&.\,\,\,\,.\,\,\,\,.\,\,\,\,.\,\,\,\,.&.\,\,\,\,.\,\,\,\,.\,\,\,\,.\,\,\,\,.&.\,\,\,\,.\,\,\,\,.\,\,\,\,.\,\,\,\,.&
7507$\pm$1456& 98.35$\pm$13.38& 12.85$\pm$4.31&0 &2$^{ab}$ \\
PG1411+442&124.3$^{+61.0}_{-61.7}$ & 443$\pm$146 & 3932$\pm$786&
11.34$\pm$2.27& 4.46$\pm$1.34& 5149$\pm$1029& 20.27$\pm$4.05&
2.53$\pm$0.76&1$^{a}$&1 \\ PG1426+015& 95.0$^{+29.9}_{-37.1}$ &
1298$\pm$385
&.\,\,\,\,.\,\,\,\,.\,\,\,\,.\,\,\,\,.&.\,\,\,\,.\,\,\,\,.\,\,\,\,.\,\,\,\,.&.\,\,\,\,.\,\,\,\,.\,\,\,\,.\,\,\,\,.&
7063$\pm$1569& 117.56$\pm$17.67& 30.78$\pm$6.79&0 &4 \\ PG1613+658&
40.1$^{+15.0}_{-15.2}$ & 279$\pm$129
&.\,\,\,\,.\,\,\,\,.\,\,\,\,.\,\,\,\,.&.\,\,\,\,.\,\,\,\,.\,\,\,\,.\,\,\,\,.&.\,\,\,\,.\,\,\,\,.\,\,\,\,.\,\,\,\,.&
6890$\pm$533& 83.64$\pm$11.29& 17.38$\pm$3.72&0 &7 \\ PG1617+175&
71.5$^{+29.6}_{-33.7}$ & 594$\pm$138 & 4896$\pm$955& 18.86$\pm$0.50&
10.23$\pm$1.38& 7452$\pm$1589& 61.98$\pm$28.47& 17.87$\pm$3.95&2 &2 \\
PG2130+099&158.1$^{+29.8}_{-18.7}$ & 457$\pm$55 & 3021$\pm$154&
10.16$\pm$1.45& 5.63$\pm$0.67& 4074$\pm$54& 52.21$\pm$6.18&
11.14$\pm$2.20&4 &3 \\

 \hline
 \hline

 \end{tabular}
\begin{list}{}{}
 \item Notes: $^{a}\,$ spectra from $HST$; $^{ab}\,$ one spectrum from $IUE$ and
 the other from $HST$; spectra from $IUE$ if not labeled.
\end{list}
\end{small}
 \end{table}
 \end {center}

\label{lastpage}

\end{document}